\documentclass[aps,twocolumn,showpacs,preprintnumbers]{revtex4}
\usepackage{psfig}

\usepackage{graphicx}
\usepackage{dcolumn}
\usepackage{bm}
\usepackage{graphics}
\usepackage{amssymb}

\begin{document}
\title{A New Window Onto Quantum Chaos}
\author{John Evans}
\email{evans@physics.ucf.edu}
\author{Fredrick Michael}
\email{fnm@physics.ucf.edu}
\affiliation{Department of Physics, University of Central Florida, Orlando, FL
32816-2385}

\date{August 7, 2002}

\begin{abstract}
In this article the statistical properties of symmetrical random matrices
whose elements are drawn from a $q$-parameterized non-extensive statistics
power-law distribution are investigated. In the limit as $q\rightarrow 1$
the well known Gaussian orthogonal ensemble (GOE) results are recovered. The
relevant level spacing distribution is derived and one obtains a suitably
generalized non-extensive Wigner distribution which depends on the value of
the tunable non-extensivity parameter $q$. This non-extensive Wigner
distribution can be seen to be a one-parameter level-spacing distribution
that allows one to interpolate between chaotic and nearly integrable regimes.
\end{abstract}
\maketitle

\bigskip
Over the past two decades random matrix theory has proven to be a useful
tool in the study of quantum chaos. Universal signatures of quantum chaos
were found for systems whose corresponding classical Hamiltonians exhibit
chaotic behavior, and they are reviewed in \cite{Gut,haake1,stockmann1}. The
level spacing distributions derived from ensembles of random matrices \cite
{Porter,mehta1} have been shown to describe closely the statistics of
fluctuations in the energy spectra of some chaotic systems \cite{Bohigas}.
This concept has also been extended to scattering \cite{Blumel,John},
dissipative systems \cite{Grobe},inter-series mixing in helium \cite{puttner1}
and in general the analysis of systems with mixed regular and chaotic
phase space regions \cite{reichl1}. 

In the generic case of mixed phase space, the Hamiltonian can be written as $%
H=H_{o}+\epsilon H_{1}$, with $H_{o}$ an integrable Hamiltonian and $H_{1}$
a non-integrable Hamiltonian. Regular and chaotic
regions coexist in the mixed phase space and no universal laws have been
shown to hold. Therefore level spacing distributions between neighboring
eigenvalues have been suggested \cite{brody1,Izrailev,berry1,stockmann1} that are
composed of the distributions describing the regular $H_{o}$ and chaotic $%
H_{1}$ regions separately. The regular regions have classically integrable
Hamiltonians $H_{o}$ whose eigenvalues are uniformly distributed and
therefore the nearest-neighbor spacing statistics are described by Poisson
distributions \cite{brody1}. The eigenvalues of the non-integrable
Hamiltonians $H_{1}$ in the chaotic regions on the other hand are assumed to
have an ergodic phase space region and are thus taken to be Gaussian
distributed. The nearest-neighbor statistics for the chaotic regions then
follow the Wigner-surmised level spacing statistics. These composite
distributions then contain one or more adjustable parameters that allow for
an interpolation between the Poisson and the Wigner level statistics.

The individual regions have Hamiltonians $H_{o},H_{1}$ whose eigenvalues are
assumed to be independent. Thus an assumption of statistical independence of
the random matrix elements (the eigenvalues) is inherent to the theory also
as a consequence of a semi-classical approximation. This assumption is
admittedly invalid in even the simplest quantum case \cite{berry1} if
tunneling between the different regions is not neglected.
 
In this letter the level spacing distribution of a generic $H$ is obtained
from statistically dependent random matrix elements corresponding to
correlated eigenvalues. The importance of using non-extensive statistic
is that the nearest-neighbor level spacing distribution allows one to interpolate from the nearly integrable to the chaotic regime as the non-extensive 
parameter $q$ is varied. The Hamiltonian matrix of $H$ to be considered 
is an $N\times N$ real symmetric matrix, and is invariant under 
orthogonal transformations. 

The extensive Gaussian random matrix theory can 
be generalized by examining the non-extensive, or $q$-parameterized entropy \cite{tsallis1,wang1}.
For systems with statistically dependent (say, Hamiltonian matrix elements)
variables the joint probability decomposition is
\begin{equation}
P(H_{i},H_{j})=P(H_{i}\mid H_{j})P(H_{j}),
\end{equation}
which gives the pseudo-additive entropy

\begin{eqnarray}
S_{q}(H_{i},H_{j})=S_{q}(H_{j})+S_{q}(H_{i}\mid H_{j})  \nonumber \\
+(1-q)S_{q}(H_{j})S_{q}(H_{i}\mid H_{j}),  \nonumber \\
S_{q}=-ln_{q}P=-\frac{P^{1-q}-1}{1-q}.
\end{eqnarray}

It is known \cite{rajagopal1} that the Tsallis entropy satisfies this
condition, and the resulting probability will be of the power-law form. The $%
q$-logarithm is $ln_{q}X=-\frac{1-X^{1-q}}{(q-1)}$. In the limit as $%
q\rightarrow 1$ the usual form of the natural logarithm and thus the
extensive statistics and its exponential (Gaussian) distributions is
recovered.

The entropy to be maximized given the constraints is then 
\begin{eqnarray}
\left\langle S\right\rangle _{q} &=&-\frac{1-\int P_{N}^{q}(H)dH}{(q-1)}%
\text{ },  \nonumber \\
\left\langle Tr(H^{2})\right\rangle _{q} &=&\int Tr\left( H^{2}\right) \text{
}P_{N}^{q}(H)dH  \nonumber \\
&=&\sigma _{q}^{2},
\end{eqnarray}
and which is subject to the extra normalization condition

\begin{center}
\begin{equation}
\int P_{N}(H)dH=1.
\end{equation}
\end{center}

Then maximizing $\left\langle S\right\rangle _{q}$ with the above
constraints, yields the least biased probability density
or non-extensive distribution function 
\begin{equation}
P_{N}(H)=A_{N}\text{ }\left( 1+\beta (q-1)Tr(H^{2})\right) ^{\frac{-1}{q-1}},
\end{equation}
where $A_{N}$ is a normalization constant and

\begin{equation}
Tr(H^{2})=\sum_{1\leq i\leq N}H_{ii}^{2}+2\sum_{1\leq i\leq j\leq
N}H_{ij}^{2}.
\end{equation}
This is then the Tsallis power-law form of the probability density for the $%
N\times N$ random matrix elements. This nonextensive distribution then is
use to obtain the correlated distribution function for the eigenenergies
assuming that the given Hamiltonian is symmetric, real and can be
diagonalized by means of an orthogonal transformation. The correlated
distribution is given as

\begin{eqnarray}
P_{N}(E_{1},...,E_{N})=  \nonumber \\
A_{N}\prod_{j>i}\left| E_{i}-E_{j}\right| \left( 1+\beta
(q-1)\sum_{i}E_{i}^{2}\right) ^{\frac{-1}{q-1}}.
\end{eqnarray}
Consider a special case of non-extensive ensembles of $2\times 2$ matrices.
The nearest neighbor spacing distribution $p(s)$ is obtained from the
correlated energy distribution function $P(E_{1},E_{2})$ by

\begin{eqnarray}
p(s)=\int_{-\infty }^{\infty }dE_{1}\int_{-\infty }^{\infty
}dE_{2}P(E_{1},E_{2})\delta (s-\left| E_{1}-E_{2}\right| )  \nonumber \\
p(s)=A\int_{-\infty }^{\infty }dE_{1}\int_{-\infty }^{\infty }dE_{2}\left|
E_{1}-E_{2}\right|  \nonumber \\
\times\left ( 1+\beta (q-1)\sum_{i}E_{i}^{2}\right) ^{\frac{-1}{q-1}}\delta
(s-\left| E_{1}-E_{2}\right| ).
\end{eqnarray}
The constants $A$ and $\beta $ are fixed by the two normalization conditions$\int p(s)ds=1$,
$\int sp(s)ds=1$.

Integrating Eq.7. with the above conditions yields the non-extensive
nearest neighbor spacing distribution. 
\begin{equation}
p(s)=\frac{(5-3q)\beta }{2}s\left( 1+\beta (q-1)s^{2}/2\right) ^{\frac{-1}{%
q-1}+\frac{1}{2}},
\end{equation}
where 
\begin{equation}
\beta (q)=\frac{\pi }{2}\frac{1}{q-1}\left( \frac{\Gamma (\frac{1}{q-1}-2)}{%
\Gamma (\frac{1}{q-1}-\frac{3}{2})}\right) ^{2},
\end{equation}
and $1<q<\frac{3}{2}$.

The behavior of the non-extensive nearest neighbor spacing distribution obtained by plotting
its values for a range of the level spacing $s$, given the `inverse
variance' $\beta $ and the non-extensivity parameter $q$. In Fig.1. the $q$
dependence of $\beta $ is plotted for values of $q$ between $1<q<1.5$. In
Fig.2. the Poisson (solid), extensive (long-short dashing) and non-extensive
Wigner (short dashing) distributions are plotted for a low non-extensivity
parameter $q$ value of $q=1.01$. The non-extensive distribution is nearly
superimposed on the extensive Wigner distribution as is expected for $q->1$.
In Fig.3. the Poisson (solid), extensive (long-short dashing) and
non-extensive Wigner (short dashing) are plotted for a high value of the
non-extensivity parameter $q=1.38$. Here the distribution is greatly shifted
and approaches the Poisson level statistics distribution.
\begin{center}
\begin{figure}[tbp]
\psfig{file=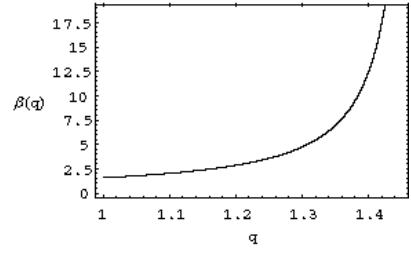,width=150pt}
\caption{$\protect\beta$ Vs. $q$}
\label{betafig}
\end{figure}
\end{center}

\begin{center}
\begin{figure}[tbp]
\psfig{file=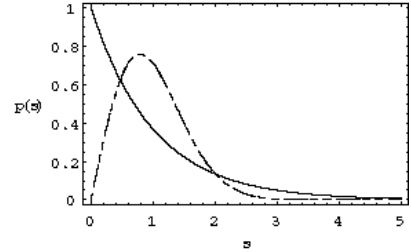,width=150pt}
\caption{$P\left( s\right)$ Vs. $s$, $q=1.01$. The non-extensive and
extensive Wigner distributions are nearly super-imposed. The Poisson
distribution is plotted using a solid line.}
\label{lowq}
\end{figure}
\end{center}
\begin{center}
\begin{figure}[tbp]
\psfig{file=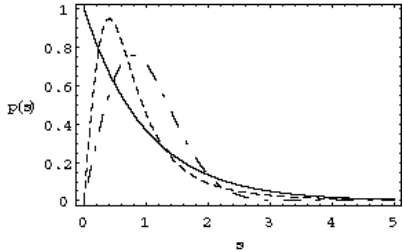,width=150pt}
\caption{$P\left( s\right)$ Vs. $s$, $q=1.38$. The extensive
distribution has been plotted with long-short dashing. The non-extensive
Wigner distribution is plotted with the short dashes. The Poisson
distribution is plotted using a solid line.}
\label{highq}
\end{figure}
\end{center}
In this letter Gaussian distributed $N\times N$ random Hamiltonian matrix
elements is generalized to the case of the non-extensive statistics and the
resultant power-law distributions. A derivation of the subsequent level
spacing statistical distribution, the non-extensive Wigner distribution, is
given. This derivation is obtained from maximizing the non-extensive entropy
of Tsallis of the $N\times N$ symmetrical random matrix elements. The
resultant of nonextensive level-spacing distribution is $q$-parameterized
and the nearest neighbor spacing distribution varies from near integrable to
chaotic regimes as the non-extesive parameter $q$ is varied. The major importance
is the possibility of using non-extensive statistics to shed light on the
quantum signatures of classically mixed systems. In future work
it will be interesting to apply these results to Hamiltonians of mixed
systems between regular and chaotic regimes where deviations from the Wigner statistics become pronounced.

\bigskip One of the authors, F. Michael, wishes to acknowledge support from the NSF
through grant number DMR99-72683.

\end{document}